\documentclass{PoS}

\title{VERITAS long term monitoring of Gamma-Ray emission from the BL Lacertae object }

\ShortTitle{VERITAS long term monitoring of Gamma-Ray emission from the BL Lacertae object }

\author{\speaker{A. U. Aeysekara for the VERITAS Collaboration}\\
        University of Utah\\
        E-mail: \email{a.abeysekara@utah.edu}}

\abstract{BL Lacertae is the prototype of the blazar subclass known as BL Lac type objects.  BL Lacertae object itself is a low-frequency-peaked BL Lac(LBL). Very high energy (VHE) gamma ray emission from this source was discovered in 2005 by MAGIC observatory while the source was at a flaring state. Since then, VHE gamma rays from this source has been detected several times. However, all of those times the source was in a high activity state.  
Former studies suggest several non-thermal zones emitting in gamma-rays, then gamma-ray flare should be composed of a convolution.
Observing the BL Lacertae object at quiescent states and active states is the key to disentangle these two components. VERITAS is monitoring the BL Lacertae object since 2011. The archival data set includes observations during flaring and quiescent states. This presentation reports on the preliminary results of the VERITAS observation between January 2013 - December 2015, and simultaneous multiwavelength observations.}

\FullConference{35th International Cosmic Ray Conference --- ICRC2017\\
		10--20 July, 2017\\
		Bexco, Busan, Korea}

\begin{document}

\section{Introduction}
BL Lacertae, also known as TeV J2202+422, 1ES 2200+420, and 3FGL J2202.7+4217, is the prototype of the blazar subclass known as BL Lac type objects. 
BL Lac objects are characterized by their highly variable non-thermal spectra, lack of prominent emission lines, and significant optical polarization that varies with time.
Literature reports the flux measurements of the BL Lacertae during its active states.
However, understanding the properties during quiescent states is also important to fully understand the physical processes of the source.

Since 2011 VERITAS regularly observed the BL Lacertae, as well as following up of active states detected by other observatories such as \textit{Fermi}-LAT.
Regular monitoring observations were scheduled for roughly four times per month with each observation of one hour exposure.  
However, the schedule can be changed due weather conditions.
This paper reports on the observations performed between 2013 January and 2015 December.

\section{VERITAS Observations}

Very Energetic Radiation Imaging Telescope Array System, VERITAS, is an array of four imaging atmospheric Cherenkov telescopes (IACTs) located at the Fred Lawrence Whipple Observatory in southern Arizona ($31^\circ ~ 40^\prime$ N, $110^\circ 57^\prime$ W, 1.3 km a.s.l.).
The array is sensitive to gamma rays in the energy range from $\sim 85$ GeV to $\sim30$ TeV, and has the sensitivity to detect a point-like source at five standard deviations (at $5 \sigma$) with a brightness of 1\% of the Crab Nebula flux with an exposure of  $<25$ hours.
The energy of gamma rays can be measured with a resolution of $15-25\%$ and the angular resolution is better than 0.1 degree at 1TeV.

\begin{figure}[ht]
\centering
 \includegraphics[width=0.5\textwidth]{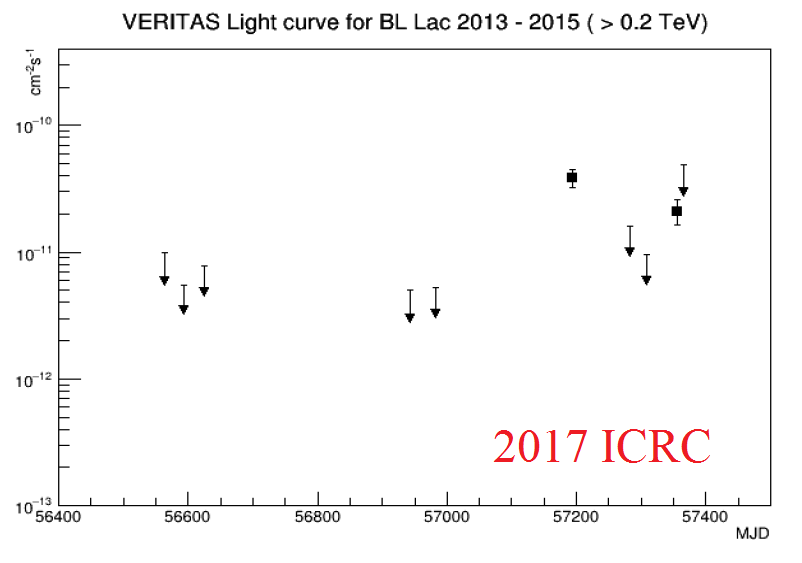}
\caption{VERITAS light curve of the BL Lacertae between 2013 January and 2015 December.}
\label{Fig:VERITASLightCurve}
\end{figure}

During the 24 months time period between January 2013 and December 2015 the BL Lacertae were observed for 1350 hours.
The entire data set has been analyzed with reflected-region background model \cite{AharonianRefflected}, 
and minimum two telescope triggering criterion described in \cite{Holder2006}. 
The light curve between 2013 and 2015 in monthly bins is shown in Figure~\ref{Fig:VERITASLightCurve}.
For the two bins centered at 2015/06/21 (Modified  Julian Date 57194), and 2015/11/30 (Modified  Julian Date 57356), gamma-ray like events were detected with a significance greater than $5 \sigma$.
Flux measurements and the detection significance for these two data points are shown in Table \ref{Table:Detections}.
Detection significance is lower than $5\sigma$ for all the other bins.
Therefore, 99\% flux upper limits for these bins were shown in Figure~\ref{Fig:VERITASLightCurve}.

\begin{table}[h]
\centering
\begin{tabular}{  | c | c | c |}
  \hline
  Bin Center & Significance & Integral flux  \\
  Date	 (Modified Julian Date)      & $\sigma$    &        $\times 10^{-11}$ ph cm$^{-2}$s$^{-1}$\\
  \hline
2015/06/21 (57194) & 7.1 & 3.85 \\
2015/11/30 (57356) & 6.9 & 2.11 \\
  \hline  
\end{tabular}

\caption{Summary of the VERITAS measurements for two flaring states. 
The first column shows the dates correspond with the center of the bin. 
The second column is the detection significance.
The third column shows the integrated flux energies greater than 200 GeV.}
\label{Table:Detections}
\end{table}

As it can be seen in  Figure~\ref{Fig:VERITASLightCurve} seven out of eight upper limits are lower than the measured flux of the two bins with the detection significance greater than $5\sigma$.
This implies that during those seven observations the BL Lacertae was in a state with fluxes less than the detected flaring states. 
Therefore, those eight bins were combined to obtain the average flux during low-states.
However, the combined flux were still less than $5\sigma$, and the derived 99\% flux upper limit is 2.50 $ \times 10^{-12}$ .

\section{\textit{Fermi}-LAT Observations During Flaring States}

In order to obtain a quasi-simultaneous GeV gamma-ray flux for the BL Lacertae during the two flaring states, 
the \textit{Fermi}-LAT data set has been analyzed from 2015/06/20 00:00 to 2015/06/20 23:59 and from 2015/11/30 00:00 to 2015/11/30 23:59, overlapping the time of the VERITAS detection of flares.
The data set has been analyzed using the Pass 8 \textit{Fermi}-LAT analysis tools.
This analysis yields a detection of the BL Lacertae with a TS = 82.8 on 2015/06/20, and with a TS = 13.3 on 2015/11/30.
Integrated fluxes between 1 and 100 GeV are $(1.2 \pm 0.4) \times 10^{-7}$ ph cm$^{-2}$ s$^{-1}$ and $(3 \pm 2) \times 10^{-8}$  ph cm$^{-2}$s$^{-1}$, respectively. 

\section{Discussion and Conclusions}
The BL Lacertae was mostly in a low-flux state during the time period between 2013 January and 2015 December, except for two occurrences.
VERITAS exposure during the time period between 2013 and 2015 was not deep enough to detect the BL Lacertae's flux during the low-flux states.
However, this paper does not report the VERITAS full data set on the BL Lacertae.
Results with the larger data sample will be published in future.

The four year averaged flux between 1 GeV and 100 GeV reported in \textit{Fermi}-LAT 3FGL source catalog is  $(1.65 \pm 0.03) \times 10^{-8}$ ph cm$^{-2}$ s$^{-1}$.
Measured \textit{Fermi}-LAT flux coincident with the VERITAS detected flare on 2015/06/20 is in a significantly elevated state.
However, the  \textit{Fermi}-LAT flux coincident with the VERITAS detected flare on 2015/11/30 is consistent with the four year averaged flux reported in the 3FGL catalog.

\section{Acknowledgments}
This research is supported by grants from the U.S. Department of Energy Office of Science, the U.S. National Science Foundation and the Smithsonian Institution, and by NSERC in Canada. We acknowledge the excellent work of the technical support staff at the Fred Lawrence Whipple Observatory and at the collaborating institutions in the construction and operation of the instrument.

\end{document}